\documentstyle[psfig]{mn}

\def\etal{{\it et al.\ }}
\def\eg{{\it e.g.\ }}

\def\spose#1{\hbox to 0pt{#1\hss}}
\def\approxlt{\mathrel{\spose{\lower 3pt\hbox{$\sim$}}
        \raise 2.0pt\hbox{$<$}}}
\def\approxgt{\mathrel{\spose{\lower 3pt\hbox{$\sim$}}
        \raise 2.0pt\hbox{$>$}}}
\def\approxpropto{\mathrel{\spose{\lower 3pt\hbox{$\sim$}}
        \raise 2.0pt\hbox{$\propto$}}}
\mathchardef\twiddle="2218

\def\multleft#1{\hbox to size{\vbox {\halign {\lft{##}\cr #1}}\hfill}\par}
\def\multright#1{\hbox to size{\vbox {\halign {\rt{##}\cr #1}}\hfill}\par}

\def\today{\ifcase\month\or January\or February\or March\or April\or May\or
      June\or July\or August\or September\or October\or November\or December\fi
      \space\number\day, \number\year}
\def\<{\thinspace}

\def\erg{{\rm\thinspace erg}}

\def\km{{\rm\thinspace km}}

\def\Mpc{{\rm\thinspace Mpc}}

\def\s{{\rm\thinspace s}}

\def\ergps{\hbox{$\erg\s^{-1}\,$}}

\def\kmps{\hbox{$\km\s^{-1}\,$}}

\def\kmpspMpc{\hbox{$\kmps\Mpc^{-1}$}}

\def\OM{\Omega_{\rm m}}
\def\OL{\Omega_{\Lambda}}
\def\OK{\Omega_{\rm K}}

\def\OX{\Omega_{\rm X}}

\def\OBH{\Omega_{\rm b}h^2}

\title[Constraints on dark energy from the largest relaxed galaxy clusters]
{Constraints on dark energy from Chandra observations of 
the largest relaxed galaxy clusters}

\author[S.W. Allen et al.]
{\parbox[]{6.in} {S.W. Allen$^1$, R.W. Schmidt$^2$, H. Ebeling$^3$, A.C. Fabian$^1$ and 
L. van Speybroeck$^4$  \\
\footnotesize
1. Institute of Astronomy, Madingley Road, Cambridge CB3 0HA \\
2. Institut f\"ur Physik, Universit\"at Potsdam, Am Neuen 
Palais 10, 14469 Potsdam, Germany \\
3. Institute for Astronomy, 2680 Woodlawn Drive, Honolulu, Hawaii 96822, USA \\
4. Harvard-Smithsonian Center for Astrophysics, 60 Garden Street, Cambridge MA 02138, USA \\
 }}
\begin{document}
\maketitle
\begin{abstract}
We present constraints on the mean dark energy density, $\OX$ and dark
energy equation of state parameter, $w_{\rm X}$, based on Chandra
measurements of the X-ray gas mass fraction in 26 X-ray luminous,
dynamically relaxed galaxy clusters spanning the redshift range $0.07<
z<0.9$. Under the assumption that the X-ray gas mass fraction measured
within $r_{2500}$ is constant with redshift and using only weak priors
on the Hubble constant and mean baryon density of the Universe, we
obtain a clear detection of the effects of dark energy on the
distances to the clusters, confirming (at comparable significance)
previous results from Type Ia supernovae studies. For a standard
$\Lambda$CDM cosmology with the curvature $\OK$ included as a free
parameter, we find $\OL=0.94^{+0.21}_{-0.23}$ (68 per cent confidence
limits). We also examine extended XCDM dark energy models.  Combining
the Chandra data with independent constraints from cosmic microwave
background experiments, we find $\Omega_{\rm X}=0.75\pm0.04$,
$\Omega_{\rm m}=0.26^{+0.06}_{-0.04}$ and $w_{\rm X}=-1.26\pm0.24$.
Imposing the prior constraint $w_{\rm X}>-1$, the same data require
$w_{\rm X}<-0.7$ at 95 per cent confidence. Similar results on the
mean matter density and dark energy equation of state parameter,
$\Omega_{\rm m}=0.24\pm0.04$ and $w_{\rm X}=-1.20^{+0.24}_{-0.28}$,
are obtained by replacing the CMB data with standard priors on the
Hubble constant and mean baryon density and assuming a flat geometry.
\end{abstract}

\begin{keywords}
X-rays: galaxies: clusters -- cosmological parameters -- dark matter -- 
cosmic microwave -- gravitational lensing 
\end{keywords}

\section{Introduction}

The matter content of the largest clusters of galaxies is thought to
provide an almost fair sample of the matter content of the Universe
(\eg White \etal 1993; Eke \etal 1998).  The observed ratio of
baryonic-to-total mass in such clusters should therefore closely match
the ratio of the cosmological parameters $\Omega_{\rm b}/\Omega_{\rm
m}$, where $\Omega_{\rm b}$ and $\Omega_{\rm m}$ are the mean baryon
and total mass densities of the Universe in units of the critical
density. The combination of robust measurements of the baryonic mass
fraction in the largest galaxy clusters with accurate determinations
of $\Omega_{\rm b}h^2$ from cosmic nucleosynthesis calculations
(constrained by the observed abundances of light elements at high
redshifts) and/or cosmic microwave background (CMB) experiments, with
a reliable measurement of the Hubble constant, $H_0$, can therefore be
used to determine $\Omega_{\rm m}$.

This method for measuring $\Omega_{\rm m}$, which is particularly
attractive in terms of the simplicity of its underlying assumptions,
was first highlighted by White \& Frenk (1991) and subsequently
employed by a number of groups (\eg Fabian 1991, White \etal 1993,
David, Jones \& Forman 1995; White \& Fabian 1995; Evrard 1997; Mohr,
Mathiesen \& Evrard 1999; Ettori \& Fabian 1999; Roussel, Sadat \&
Blanchard 2000; Allen, Schmidt \& Fabian 2002a; Allen \etal 2003;
Ettori, Tozzi \& Rosati 2003; Sanderson \& Ponman 2003; Lin, Mohr \&
Stanford 2003). In general, these studies have found $\Omega_{\rm m}<
1$ at high significance, with recent work favouring $\Omega_{\rm
m}\sim0.3\,h_{70}^{-0.5}$.

Sasaki (1996) and Pen (1997) were the first to describe how
measurements of the apparent redshift dependence of the baryonic mass
fraction could also, in principle, be used to constrain the geometry
and, therefore, dark energy density of the Universe.  The geometrical
constraint arises from the dependence of the measured baryonic mass
fraction values on the assumed angular diameter distances to the
clusters. Although theory and cosmological simulations suggest that
the baryonic mass fraction in the largest clusters should be invariant
with redshift (\eg Eke \etal 1998), this will only $appear$ to be the
case when the reference cosmology used in making the baryonic mass
fraction measurements matches the true, underlying cosmology.  The
first successful application of such a test was carried out by Allen,
Schmidt \& Fabian (2002a; hereafter ASF02; see also Allen \etal 2003a)
using a small sample of X-ray luminous, dynamically relaxed clusters
with precise mass measurements, spanning the redshift range
$0.1<z<0.5$. These authors found $\Omega_{\rm m}=0.30\pm0.04$ and
$\Omega_{\Lambda}=0.95^{+0.48}_{-0.72}$.  A similar analysis was later
carried out by Ettori, Tozzi \& Rosati (2003) who obtained consistent
results using a larger cluster sample that extended to higher
redshift, but which included clusters with a wider range of dynamical
states.

The baryonic mass content of galaxy clusters is dominated by the X-ray
emitting intracluster gas, the mass of which exceeds the mass of
optically luminous material by a factor $\sim 6$ (\eg White \etal
1993; Fukugita, Hogan \& Peebles 1998; other mass components in
clusters are expected to make only very small contributions to the
total baryon budget). Since the emissivity of the X-ray emitting gas
is proportional to the square of its density, the gas mass profile in
a cluster can be precisely determined from X-ray data. Measuring the
$total$ mass profile is more difficult, however, and requires both a
direct measurement of the X-ray gas temperature profile and the
assumption of hydrostatic equilibrium in the gas.  Measurements of the
temperature profiles in intermediate-to-high redshift galaxy clusters
have only become possible following the launch of the Chandra X-ray
Observatory. The exquisite spatial resolution of Chandra makes
measuring the temperature profiles of even distant clusters a
relatively straightforward task, given sufficient exposure time. The
use of the hydrostatic assumption in making the mass and baryonic mass
fraction measurements is more problematic, however, and requires a
restriction to dynamically relaxed systems when carrying out a
cosmological test similar to that described here.

In this paper we present a significant extension to the ASF02 work.
The cluster sample is significantly larger and includes 26 X-ray
luminous, dynamically relaxed systems spanning the redshift range
$0.07<z<0.9$. As well as enhancing the sample, we have also expanded
the analysis. We now include a bias factor, $b$, in the X-ray analysis
(see also Allen, Schmidt \& Bridle 2003), motivated by gas-dynamical
simulations, that accounts for the (relatively small amount of)
baryonic material expelled from such clusters as they form. We also
examine XCDM dark energy models in which the equation of state
parameter, $w_{\rm X}$, is allowed to take any constant
value. Finally, as well as results based on the Chandra data using
simple priors on $\Omega_{\rm b}h^2$, $h$ and $b$, we also present
results from the combination of Chandra and CMB data. This latter
combination is shown to be particularly powerful in constraining the
overall dark energy density.
 
As in ASF02, we report measurements of the X-ray gas mass fraction for 
two reference cosmologies: an SCDM cosmology with $h = H_0/100$\kmpspMpc 
$= 0.5$, $\Omega_{\rm m} = 1$ and $\Omega_\Lambda = 0$, and a 
$\Lambda$CDM cosmology with $h=0.7$, $\Omega_{\rm m} = 0.3$ and 
$\Omega_\Lambda = 0.7$.

\section{Observations and data reduction}

\begin{table*}
\begin{center}
\caption{Summary of the Chandra observations. The columns list the 
target name, observation date, detector used, observation mode 
and net exposure after all cleaning and screening processes were
applied. The targets are listed in RA order.}\label{table:obs}
\vskip 0 truein
\begin{tabular}{ c c c c c c }
&&&&&  \\
                    & ~ &      Date      & Detector & Mode & Exposure (ks) \\
\hline
MACSJ0242.6-2132    & ~ &   2002 Feb 07  & ACIS-I & VFAINT & 10.2 \\
Abell 383(1)        & ~ &   2000 Nov 16  & ACIS-S & VFAINT & 18.0  \\
Abell 383(2)        & ~ &   2000 Nov 16  & ACIS-I & VFAINT & 17.2  \\
MACSJ0329.7-0212    & ~ &   2002 Dec 12  & ACIS-I & VFAINT & 16.8 \\
Abell 478           & ~ &   2001 Jan 27  & ACIS-S & FAINT  & 39.9  \\
MACSJ0429.6-0253    & ~ &   2002 Feb 07  & ACIS-I & VFAINT & 19.1 \\
RXJ0439.0+0520      & ~ &   2000 Aug 29  & ACIS-I & FAINT  & 7.6  \\
MACSJ0744.9+3927(1) & ~ &   2001 Nov 12  & ACIS-I & VFAINT & 17.1 \\
MACSJ0744.9+3927(2) & ~ &   2003 Jan 04  & ACIS-I & VFAINT & 14.5 \\
Abell 611           & ~ &   2001 Nov 03  & ACIS-S & VFAINT & 34.5  \\
4C55                & ~ &   2004 Jan 03  & ACIS-S & VFAINT & 77.8  \\
MACSJ0947.2+7623    & ~ &   2000 Oct 20  & ACIS-I & VFAINT & 9.6 \\
Abell 963           & ~ &   2000 Oct 11  & ACIS-S & FAINT  & 34.8  \\
MS1137.5+6625       & ~ &   1999 Sep 30  & ACIS-I & VFAINT & 103.8 \\
Abell 1413          & ~ &   2001 May 16  & ACIS-I & VFAINT & 8.1   \\
ClJ1226.9+3332      & ~ &   2003 Jan 27  & ACIS-I & VFAINT & 25.7 \\
MACSJ1311.0-0311    & ~ &   2002 Dec 15  & ACIS-I & VFAINT & 12.0 \\
RXJ1347.5-1145(1)   & ~ &   2000 Mar 03  & ACIS-S & VFAINT & 8.6   \\
RXJ1347.5-1145(2)   & ~ &   2000 Apr 29  & ACIS-S & FAINT  & 10.0  \\
RXJ1347.5-1145(3)   & ~ &   2003 Sep 03  & ACIS-I & VFAINT & 49.3  \\
Abell 1835(1)       & ~ &   1999 Dec 11  & ACIS-S & FAINT  & 18.0  \\
Abell 1835(2)       & ~ &   2000 Apr 29  & ACIS-S & FAINT  & 10.3  \\
3C295               & ~ &   1999 Aug 30  & ACIS-S & FAINT  & 17.0  \\
MACSJ1423.8+2404    & ~ &   2003 Aug 18  & ACIS-S & VFAINT & 112.5 \\
Abell 2029          & ~ &   2000 Apr 12  & ACIS-S & FAINT  & 19.2  \\
MACSJ1532.9+3021(1) & ~ &   2001 Aug 26  & ACIS-S & VFAINT & 9.4 \\
MACSJ1532.9+3021(2) & ~ &   2001 Sep 06  & ACIS-I & VFAINT & 9.2 \\
MACSJ1621.6+3810    & ~ &   2002 Oct 18  & ACIS-I & VFAINT & 7.9 \\
MACSJ1720.3+3536    & ~ &   2002 Nov 03  & ACIS-I & VFAINT & 16.6 \\
MACSJ1931.8-2635    & ~ &   2002 Oct 20  & ACIS-I & VFAINT & 12.2 \\
MS2137.3-2353(1)    & ~ &   1999 Nov 18  & ACIS-S & VFAINT & 20.5  \\
MS2137.3-2353(2)    & ~ &   2003 Nov 18  & ACIS-S & VFAINT & 26.6  \\
MACSJ2229.8-2756    & ~ &   2002 Nov 13  & ACIS-I & VFAINT & 11.8 \\
&&&&& \\             
\hline                      
\end{tabular}
\end{center}
\end{table*}

\subsection{Sample selection}

Our sample consists of 26 galaxy clusters spanning the redshift range
$0.07<z<0.9$, with X-ray temperatures $kT\approxgt5$\,keV~and X-ray
luminosities $L_{\rm X,0.1-2.4}\approxgt 10^{45}\,h_{50}^{-2}$\ergps.
The clusters exhibit a high degree of dynamical relaxation in their
Chandra images (sharp central X-ray surface brightness peaks, regular
X-ray isophotes and minimal isophote centroid variations) and show no
evidence for a significant loss of hydrostatic equilibrium in X-ray
pressure maps and/or gravitational lensing data, where available.  The
temperature/luminosity cuts avoid complexities associated with
variations in the fraction of baryonic matter expelled from the
central regions of the clusters during their formation (Eke \etal
1998; Bialek, Evrard \& Mohr 2001; it should be possible to relax
these cuts in future work, given an improved calibration of the effect
from simulations). No quantitative morphological classification
procedure was used in the selection of the sample (although the
inclusion of such a procedure using e.g. the power ratios method of
Buote \& Tsai 1995 would be straightforward in future work). Note 
that the selection function is not required for the determination of
cosmological parameters. We simply require accurate mass measurements.

At moderate-to-high redshifts ($z>0.3$) the extension of the sample
with respect to ASF02 was achieved primarily through two Large
Programs of Chandra observations, lead by L. van Speybroeck and
H. Ebeling, of clusters in the Massive Cluster Survey (MACS; Ebeling,
Edge \& Henry 2001).  From relatively short Chandra observations of 53
individual MACS clusters, we identified 12 systems with a high degree
of dynamical relaxation (details of the cluster morphologies are
discussed by Ebeling \etal 2004, in preparation). One of the clusters,
MACSJ1423.8+2404, also has an additional deep Chandra observation
which is used here. In addition to the MACS clusters, we have also
included archival Chandra data for two other high redshift systems:
MS1137.5+6625 ($z=0.782$; Gioia \& Luppino 1994) and ClJ1226.9+3332
($z=0.892$; Ebeling \etal 2001). The central X-ray emission in these
clusters is less sharply peaked than most of the systems at lower
redshifts (although the central cooling times in the clusters are
still only $2-3 \times 10^{9}$ yrs however). However, both clusters
exhibit regular X-ray morphologies and are the most apparently
relaxed, X-ray luminous clusters that we are aware of at such high
redshifts. Additional support for the inclusion of MS1137.5+6625 in
our study comes from the agreement of the best fitting total mass
model determined from the Chandra X-ray data (presented here) and the
independent weak lensing study of Clowe \etal (2000). For an SCDM
cosmology, the Chandra data are well described by an NFW mass model
with $c=3.5^{+1.8}_{-1.5}$ and $r_{\rm
s}=185^{+225}_{-75}h^{-1}$\,kpc, implying $r_{200}=c\,r_{\rm s} =
650^{+180}_{-100}h^{-1}$\,kpc. For the same cosmology, Clowe \etal
(2000) find $c=4.2$ and $r_{200}\sim 730h^{-1}$\,kpc.  The effective
velocity dispersion corresponding to the best-fit Chandra mass model,
$\sigma=1100^{+300}_{-200}$\kmps, is also consistent with the observed
value of $\sigma=884^{+185}_{-124}$\kmps~(Donahue \etal 1999).
Although no detailed weak lensing study is available for
ClJ1226.9+3332, Maughan \etal (2004) present a temperature map from
XMM-Newton observations which supports the identification of this
system as a regular, dynamically relaxed cluster, well suited to the
present work. The observed velocity dispersion of
$\sigma=997^{+285}_{-205}$\kmps~(Maughan \etal 2004) is also
consistent with the value of $\sigma \approxgt 1000$\kmps~inferred
from the Chandra data.

At low redshifts ($z<0.3$) the extension of the sample with respect to
ASF02 was achieved via a search of the Chandra archive. Two clusters
from the ASF02 sample have been dropped from this work: PKS0745-191
($z=0.103$) was dropped due to difficulties associated with
extrapolating the observed profile beyond the restricted Chandra
ACIS-S field of view. Abell 2390 ($z=0.230$) was dropped due to
dynamical activity which is not localized and cannot easily be
excluded from the analysis. The results for both clusters listed 
in ASF02 are, however, consistent with the analysis presented here.

\subsection{Chandra observations}

\begin{table}
\begin{center}
\caption{The measured X-ray gas mass fractions at $r_{2500}$ (68 per cent 
confidence limits) for the reference SCDM and $\Lambda$CDM
cosmologies.  The results for Abell 2029 and 478 have been
extrapolated as described in the text. Redshifts for the MACS clusters
are from Ebeling \etal (2004, in preparation). A table containing the
redshift and $f_{\rm gas}$ data is available from the authors
on request.
}\label{table:fgas}
\vskip -0.1truein
\begin{tabular}{ c c c c c c c c }
&&&&  \\
\multicolumn{1}{c}{} &
\multicolumn{1}{c}{} &
\multicolumn{1}{c}{SCDM} &
\multicolumn{1}{c}{$\Lambda$CDM} \\
\multicolumn{1}{c}{} &
\multicolumn{1}{c}{} &
\multicolumn{1}{c}{{$f_{\rm gas}h_{50}^{-1.5}$}} &
\multicolumn{1}{c}{{$f_{\rm gas}h_{70}^{-1.5}$}} \\
\hline                                                                      
MACSJ0242.6-2132    &  &    $0.175\pm0.023$  & $0.130\pm0.018$   \\
Abell 383           &  &    $0.169\pm0.011$  & $0.122\pm0.009$   \\
MACSJ0329.7-0212    &  &    $0.155\pm0.019$  & $0.119\pm0.018$   \\
Abell 478           &  &    $0.184\pm0.011$  & $0.120\pm0.008$   \\
MACSJ0429.6-0253    &  &    $0.177\pm0.017$  & $0.141\pm0.015$   \\
RXJ0439.0+0520      &  &    $0.137\pm0.018$  & $0.098\pm0.013$   \\
MACSJ0744.9+3927    &  &    $0.155\pm0.018$  & $0.141\pm0.019$   \\
Abell 611           &  &    $0.149\pm0.017$  & $0.111\pm0.016$   \\
4C55                &  &    $0.163\pm0.009$  & $0.115\pm0.005$   \\
MACSJ0947.2+7623    &  &    $0.173\pm0.019$  & $0.130\pm0.016$   \\
Abell 963           &  &    $0.180\pm0.015$  & $0.128\pm0.012$   \\
MS1137.5+6625       &  &    $0.100\pm0.016$  & $0.094\pm0.014$   \\
Abell 1413          &  &    $0.167\pm0.019$  & $0.114\pm0.013$   \\
ClJ1226.9+3332      &  &    $0.114\pm0.021$  & $0.102\pm0.027$   \\
MACSJ1311.0-0311    &  &    $0.094\pm0.025$  & $0.072\pm0.022$   \\
RXJ1347.5-1145      &  &    $0.137\pm0.009$  & $0.108\pm0.009$   \\
Abell 1835          &  &    $0.164\pm0.012$  & $0.112\pm0.012$   \\
3C295               &  &    $0.129\pm0.019$  & $0.106\pm0.018$   \\
MACSJ1423.8+2404    &  &    $0.135\pm0.011$  & $0.113\pm0.008$   \\
Abell 2029          &  &    $0.189\pm0.011$  & $0.121\pm0.008$   \\
MACSJ1532.9+3021    &  &    $0.159\pm0.017$  & $0.114\pm0.017$   \\
MACSJ1621.6+3810    &  &    $0.156\pm0.034$  & $0.131\pm0.029$   \\
MACSJ1720.3+3536    &  &    $0.159\pm0.024$  & $0.123\pm0.020$   \\
MACSJ1931.8-2635    &  &    $0.189\pm0.025$  & $0.145\pm0.022$   \\
MS2137.3-2353       &  &    $0.169\pm0.010$  & $0.124\pm0.009$   \\
MACSJ2229.8-2756    &  &    $0.177\pm0.018$  & $0.139\pm0.017$   \\
&&&& \\             
\hline                                                                                                                                       
\end{tabular}                                                                                                                                
\end{center}
\end{table}

The Chandra observations were carried out using the Advanced CCD
Imaging Spectrometer (ACIS) between 1999 August 30 and 2004 Jan 03.
The standard level-1 event lists produced by the Chandra pipeline
processing were reprocessed using the $CIAO$ (version 3.0.2) software
package, including the latest gain maps and calibration products. Bad
pixels were removed and standard grade selections applied. Where
possible, the extra information available in VFAINT mode was used to
improve the rejection of cosmic ray events. For observations carried
out with the ACIS-I detector, we have used the Chandra X-ray Center (CXC)/MIT
charge transfer inefficiency correction.  Time-dependent
gain corrections were applied using A. Vikhlinin's {\it apply\_gain}
routine. The data were cleaned of periods of anomalously high
background using the same author's {\it lc\_clean} script, using the 
recommended energy ranges and bin sizes for each detector. The 
net exposure times for the individual clusters are summarized in 
Table~\ref{table:obs} and vary between 7.6 and 
112.5 ks. The total good exposure time is 825.8 ks.

The data have been analysed using the methods described by Allen \etal
(2001a, 2002) and Schmidt \etal (2001). In brief, concentric annular
spectra were extracted from the cleaned event lists, centred on the
peaks of the X-ray emission from the clusters.\footnote{For
RXJ1347-1145, the data from the southeast quadrant of the cluster were
excluded due to ongoing merger activity in that region (see Allen \etal
2002).}  The spectra were analysed using XSPEC (version 11.3: Arnaud
1996), the MEKAL plasma emission code (Kaastra \& Mewe 1993;
incorporating the Fe-L calculations of Liedhal, Osterheld \& Goldstein
1995) and the photoelectric absorption models of Balucinska-Church \&
McCammon (1992). The ACISABS model was used to account for 
time-dependent contamination along the instrument light path. 
We have allowed for the small amount of extra contamination 
present in ACIS-I observations by increasing the optical depth of 
the ACISABS model contaminant by $\Delta \tau=0.14$ at 0.67 keV 
(A. Vikhlinin, private communication).  Only data in the $0.8-7.0$ keV
energy range were used for our analysis (the exceptions being
the earliest observations of 3C295, Abell 1835 and Abell 2029 where a
wider 0.6-7.0 keV band was used).  The spectra for all annuli were
modelled simultaneously, in order to determine the deprojected X-ray
gas temperature profiles under the assumption of spherical symmetry.

For the nearer clusters ($z<0.3$), background spectra were extracted
from the blank-field data sets produced by M. Markevitch and available
from the CXC. For the more distant systems (and the first observation
of Abell 1835, which has an unusually high, but relatively constant,
background level) background spectra were extracted from appropriate,
source free regions of the target data sets. (We have confirmed that
similar results are obtained using the blank-field background data
sets throughout.)  Separate photon-weighted response matrices and
effective area files were constructed for each region studied, using
the calibration files appropriate for the period of observation. For
the ACIS-I analysis, we have decreased the quantum efficiency at
energies below 2 keV by 7 per cent from the nominal value, and then at
all energies by a further 8 per cent.  This improves the
cross-calibration between ACIS-S and ACIS-I observations of clusters
in our sample and is consistent with the recommendations of the
Chandra calibration team (A. Vikhlinin, private communication).

\subsection{Chandra $f_{\rm gas}$ measurements}

For the mass modelling, azimuthally-averaged surface brightness
profiles were constructed from background subtracted, flat-fielded
images with a $0.984\times0.984$ arcsec$^2$ pixel scale ($2\times2$
raw detector  pixels). When combined with the deprojected spectral
temperature profiles,  the surface brightness profiles can be used to
determine the X-ray gas mass profiles (to high precision) and total
mass profiles in the clusters.\footnote{The observed surface
brightness profile and parameterized mass model are
together used to predict the temperature profile of the X-ray gas. We
use the median temperature profile determined from 100 Monte-Carlo
simulations. The outermost pressure is fixed using an iterative
technique which ensures a smooth pressure gradient in these regions.
The predicted temperature profile is rebinned to the same binning as 
the spectral results and the $\chi^2$  difference between the
observed and predicted, deprojected temperature  profiles is
calculated. The parameters for the mass model are stepped through a
regular grid of values in the $r_{\rm s}$-$\sigma$ plane (see text) to
determine the best-fit values and 68 per cent confidence limits. The
gas mass profile is determined to high precision at each grid point
directly from the observed surface brightness profile and model
temperature profile. Spherical symmetry and hydrostatic equilibrium
are assumed throughout.} For this analysis, we have used an enhanced
version of the image deprojection code described by White, Jones \&
Forman (1997).

We have parameterized the cluster total mass profiles (luminous plus dark matter)
using a  Navarro, Frenk \& White (1997; hereafter NFW) model with

\begin{equation}
\rho(r) = {{\rho_{\rm c}(z) \delta_{\rm c}} \over {  ({r/r_{\rm s}}) 
\left(1+{r/r_{\rm s}} \right)^2}},
\end{equation}


\noindent where $\rho(r)$ is the mass density, 
$\rho_{\rm c}(z) = 3H(z)^2/ 8 \pi
G$ is the critical density for closure at redshift $z$, $r_{\rm s}$ is
the scale  radius, $c$ is the concentration parameter (with
$c=r_{200}/r_{\rm s}$) and  $\delta_{\rm c} = {200 c^3 / 3 \left[
{{\rm ln}(1+c)-{c/(1+c)}}\right]}$.   The normalizations of the mass
profiles may also be usefully expressed in terms of an effective 
velocity dispersion, $\sigma = \sqrt{50} r_{\rm s} c H(z)$ (with $r_{\rm s}$
in units of Mpc and $H(z)$ in \kmpspMpc). Mass models were examined
over regular $100\times100$ grids in the ($r_{\rm s}$,$\sigma$) plane.

\begin{figure}
\vspace{0.5cm}
\hbox{
\hspace{-0.5cm}\psfig{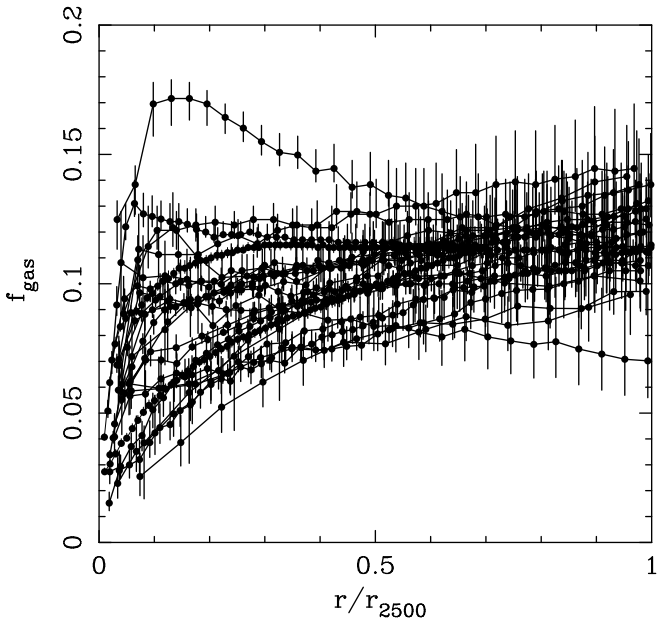}} 
\caption{The observed X-ray gas mass fraction profiles for the
26 clusters with the radial axes scaled in units of $r_{2500}$.  The
reference $\Lambda$CDM cosmology is assumed. Note that $f_{\rm gas}(r)$
is an integrated quantity and the error bars on neighbouring points
in a profile are correlated.}\label{fig:fgasr}
\end{figure}

\begin{figure*}
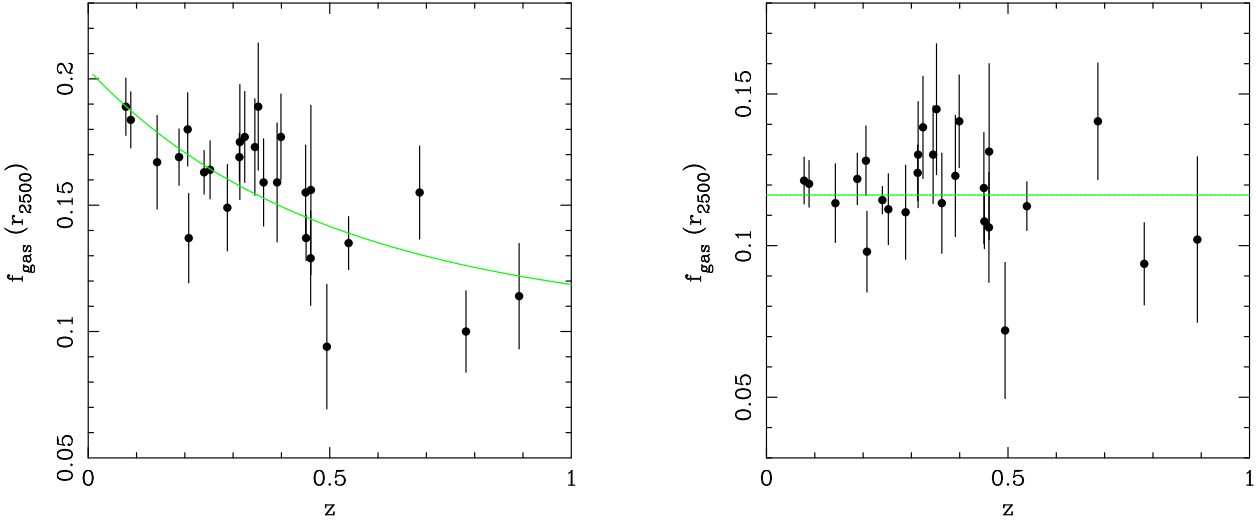

\vspace{0.2cm}
\hbox{
\hspace{0.2cm}\psfig{figure=fgas_z_scdm.ps,width=0.43 \textwidth,angle=270}
\hspace{1.3cm}\psfig{figure=fgas_z_lcdm.ps,width=0.43 \textwidth,angle=270}
}
\caption{The apparent variation of the X-ray gas mass
fraction (with root-mean-square $1\sigma$ errors) as a function of
redshift for the reference (a: left panel) SCDM and (b: right panel)
$\Lambda$CDM cosmologies. The grey curve in (a) shows the predicted
$f_{\rm gas}(z)$ behaviour for the best-fitting model cosmology
with $\Omega_{\rm m}=0.25$ and $\Omega_{\Lambda}=0.96$ (see Section
3.2). Clusters at higher redshifts appear to have lower gas mass 
fractions because the SCDM cosmology underestimates the relative 
distances to these systems. The curve in (b) shows the best-fitting 
constant value.}\label{fig:fgasz}
\end{figure*}

In determining the results on the X-ray gas mass fraction, $f_{\rm
gas}$, we have adopted a canonical radius of $r_{2500}$, within which the
mean mass density is 2500 times the critical density of the Universe
at the redshifts of the clusters. The $r_{2500}$ values are determined
directly from the Chandra data, with confidence limits calculated from
the $\chi^2$ grids and are (in general) well-matched to the outermost
radii at which reliable temperature measurements can be made from the
Chandra data.

Fig. \ref{fig:fgasr} shows the observed $f_{\rm gas}(r)$ profiles for
the 26 clusters in our sample, determined from the Chandra data using
the reference $\Lambda$CDM cosmology.  Although some variation is
present from cluster to cluster, particularly at small radii, the
profiles tend towards a universal value at $r_{2500}$. (We note that
the cluster with the most discrepant $f_{\rm gas}(r)$ profile and
highest $f_{\rm gas}$ value at small radii is MACSJ1532.9+3021, which
exhibits an unusually high ellipticity in its innermost regions. 4C55
also exhibits a high ellipticity and isophote shifts at small radii 
and has the second highest $f_{\rm gas}$ value at
$0.1r_{2500}$. Both clusters appear relaxed at larger radii, 
however, and their $f_{\rm gas}$ profiles have recovered
to the `universal' form by $r_{2500}$.)  Table
\ref{table:fgas} lists the results on the X-ray gas mass fractions at
$r_{2500}$ for the reference SCDM and $\Lambda$CDM cosmologies. Taking
the weighted mean of the $\Lambda$CDM results we obtain ${\bar f_{\rm
gas}} = 0.1173 \pm 0.0022\,h_{70}^{-1.5}$.

In calculating the total baryonic mass in the clusters, we assume 
that the optically luminous baryonic mass in galaxies scales as
$0.19h^{0.5}$ times the X-ray gas mass. This result is based on 
detailed studies of nearby and intermediate redshift clusters 
(White \etal 1993, Fukugita, Hogan \& Peebles 1998; see also 
Voevodkin \& Vikhlinin 2004) and corresponds to $\sim 16$ per cent of 
the X-ray gas mass. 
Uncertainties in this correction have a negligible impact on 
the overall error budget.
Other sources of baryonic matter in the clusters are expected to make 
very small contributions to the total mass and are ignored.

We note that the Chandra data for Abell 478 and 2029 
do not extend quite to $r_{2500}$. For these clusters, we 
measure $f_{\rm gas}$ directly at $r=0.75\,r_{2500}$ for the 
SCDM cosmology or $r=0.70\,r_{2500}$ for the $\Lambda$CDM cosmology 
and extrapolate the results to $r_{2500}$ using the median
$f_{\rm gas}(r)$ profile determined from Fig.~\ref{fig:fgasr}. This
extrapolation results in corrections to the directly measured $f_{\rm gas}$ 
values of $\sim 5$ per cent. To be conservative, we have included 
a 5 per cent systematic uncertainty in the tabulated $f_{\rm gas}$ 
measurements for Abell 478 and 2029 to allow for uncertainties in
this extrapolation. 

Fig. \ref{fig:fgasz}~shows the $f_{\rm gas}$ values as a function of
redshift for the reference SCDM and $\Lambda$CDM cosmologies. Whereas
the results for the $\Lambda$CDM cosmology are consistent with a
constant $f_{\rm gas}$ value ($\chi^2=22.7$ for 25 degrees of freedom)
the results for the reference SCDM cosmology indicate an apparent drop
in $f_{\rm gas}$ as the redshift increases. The $\chi^2=61.8$ 
obtained from a fit to the SCDM data with a constant model 
indicates that the SCDM cosmology is inconsistent 
with the expectation that $f_{\rm gas}(z)$ should be constant.

\subsection{CMB analysis}

Our analysis of CMB observations uses the WMAP temperature (TT) data
for multipoles $l<900$ (Hinshaw \etal 2003) and
temperature-polarization (TE) data for $l<450$ (Kogut \etal 2003).  To
extend the analysis to higher multipoles (smaller scales), we also
include data from the Cosmic Background Imager (CBI; Pearson \etal
2003) and Arcminute Cosmology Bolometer Array Receiver (ACBAR; Kuo
\etal 2003) for $l > 800$. The comparison of model angular power
spectra with the WMAP data employs the likelihood calculation routines
released by the WMAP team (Verde \etal 2003).

Our analysis of the CMB data uses the CosmoMC
code\footnote{{http://cosmologist.info/cosmomc/}}. This in turn
uses CAMB (Lewis, Challinor \& Lasenby 2000), which is based on
CMBFAST (Seljak \& Zaldarriaga 1996), to generate the CMB and matter
power spectrum transfer functions, and a Metropolis-Hastings Markov
Chain Monte Carlo (MCMC) algorithm to explore parameter space. We used
the covariance matrix of the parameters calculated from an initial 
set of test runs to improve sampling efficiency
(see Lewis \& Bridle 2002 for more details). 

We have fitted the data using an extended XCDM cosmological model with
eight free parameters: the physical dark matter and baryon densities
in units of the critical density, the curvature $\Omega_{\rm K}$, the
Hubble constant, the dark energy equation of state parameter, the
recombination redshift (at which the reionization fraction is a half,
assuming instantaneous reionization), the amplitude of the scalar
power spectrum and the scalar spectral index.  We also examined a
standard $\Lambda$CDM model in which we fixed $w_{\rm X}=-1$. In all
cases, we have assumed an absence of tensor components and included
uniform priors $30<H_0$\kmpspMpc$<100$ and $-4<w_{\rm X}<1$. (Tests 
in which tensor components were included with $\Lambda$CDM models 
lead to similar results on dark energy, but took much longer to 
compute.)

The analysis was carried out on the Cambridge X-ray group Linux
cluster. For each model we accumulated a total of at least $10^6$
correlated samples in 10 separate chains. We satisfied ourselves that
the chains had converged by ensuring that consistent final results
were obtained from numerous small subsets of the chains. In all cases,
we allowed a conservative burn-in period of $10^4$ samples for each
chain.

\section{Cosmological analysis}

\subsection{Dark energy models}

We have considered two separate dark energy models in our analysis: 
standard $\Lambda$CDM and the extended XCDM parameterization. Our 
definitions of the relevant quantities closely follow Peebles \& Ratra (2003) 
and we refer the reader to that work for a discussion of the underlying assumptions. 

The Friedmann equation, which relates the first time derivative of the 
scale factor of the Universe, $a$, to the total density can be 
conveniently expressed as $({\dot a}/a)^2\,$=$\,H(z)^2\,$=\,$H_0^2E(z)^2$, 
where

\vspace{0.1cm}
\begin{equation}
E(z) = \sqrt{ \OM (1+z)^3 + \OX f(z) + \OK (1+z)^2}
\label{eq:GR4}
\end{equation}
\vspace{0.1cm}

\noindent Here $\OX$ is the dark energy density and 
$f(z)$ its redshift dependence. (We have ignored the 
density contribution from radiation and relativistic matter.) 
For $\Lambda$CDM cosmologies, the dark energy density is constant 
and $f(z)=1$. Within the extended XCDM dark energy 
parametrization, the pressure is related to the density as $p_{\rm X} = 
w_{\rm X} \rho_{\rm X}$ so that for constant $w_{\rm X}$, the dark energy 
density scales as $\rho_{\rm X} \propto a^{-3(1+w_{\rm X})}$ and 
$f(z) = (1+z)^{3(1+w_{\rm X})}$. We note that for $w_{\rm X} < -1/3$ 
the dark energy makes a positive contribution to the acceleration 
of the expansion of the universe. For $w_{\rm X} < -1$ the dark 
energy density is increasing with time. 

Our analysis of Chandra $f_{\rm gas}$ data requires the angular diameter 
distances to the clusters which are defined as   

\begin{equation}
   d_{A} = \frac{c}{H_0 (1+z)\sqrt{\OK}} \, \sinh \left(\sqrt{\OK}\int_0^z
        {dz\over E(z)}\right). 
 \label{eq:DA}
\end{equation}

\subsection{Analysis of the $f_{\rm gas}$ data}\label{section:fgasmodel}

\begin{figure}
\vspace{0.5cm} \hbox{
\hspace{-0.1cm}\psfig{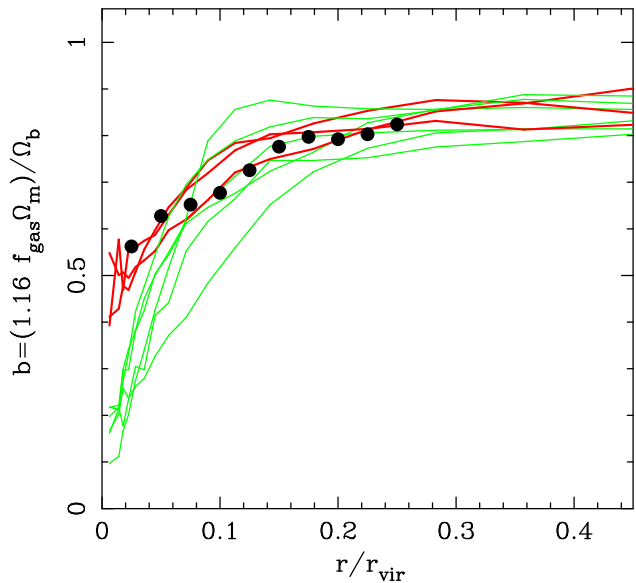}}
\caption{The X-ray bias factor, $b$ (the enclosed baryon fraction relative 
to the universal value) as a function of radius in units of the 
virial radius $r_{\rm vir}$, from the  simulations of Eke \etal (1998). 
The simulated clusters have similar masses to the systems 
studied here. The results for the three 
most dynamically relaxed clusters in the simulations are shown as darker 
curves. The solid circles mark the median
profile determined from the Chandra observations, scaled from 
Fig.~\ref{fig:fgasr} assuming $\Omega_{\rm m}=0.25$, 
$\Omega_{\rm b}=0.0413$ and $r_{2500}=0.25\,r_{\rm vir}$.
Beyond a radius $r > 0.2 r_{\rm vir}$, the simulated clusters exhibit 
consistent,  relatively flat $b$ profiles.  At
$r=0.25 r_{\rm vir}$, a radius comparable to the measurement radius
of the Chandra observations, the simulations  give
$b=0.824\pm0.033$.}\label{fig:eke}
\end{figure}

The differences between the shapes of the $f_{\rm gas}(z)$ curves
in Figs. \ref{fig:fgasz}(a) and (b) reflect the dependence 
of the measured $f_{\rm gas}(z)$ values on the angular diameter
distances to the clusters: $f_{\rm gas} \propto (1+z)^2d_{\rm A}^{1.5}$. 
Under the assumption (Section 1) that $f_{\rm gas}$ should, in reality, 
be constant with redshift, simple inspection of Fig. 
\ref{fig:fgasz} clearly favours the $\Lambda$CDM over the SCDM cosmology.

To determine constraints on the relevant cosmological parameters, 
we have fitted the $f_{\rm gas}(z)$ data in Fig. \ref{fig:fgasz}(a) with a
model that accounts for the expected apparent variation in $f_{\rm
gas}(z)$ as the underlying cosmology is varied. (We choose to 
work with the SCDM data as our reference cosmology, although 
similar results can be derived using the $\Lambda$CDM data set.)  
Note that the $f_{\rm gas}(r)$ profiles exhibit only small 
variations around $r_{2500}$ so changes 
in $r_{2500}$ as the cosmology is varied can be ignored. 
The model function fitted to the data is 

\begin{equation}
f_{\rm gas}^{\rm SCDM}(z) = \frac{ b\, \Omega_{\rm b}} {\left(1+0.19
\sqrt{h}\right) \Omega_{\rm m}} \left[ \frac{d_{\rm
A}^{\rm SCDM}(z)}{d_{\rm
A}^{\rm mod}(z)} \right]^{1.5},
\label{eq:fgas}
\end{equation}

\noindent where $d_{\rm A}^{\rm mod}(z)$ and 
$d_{\rm A}^{\rm SCDM}(z)$ are the angular diameter distances to the
clusters in the current model and reference SCDM ($h=0.5$)
cosmologies.  Note that although variations in the dark energy density
affect only the $shape$ of the $f_{\rm gas}(z)$ curve, the
normalization depends on $\Omega_{\rm m}$, $\Omega_{\rm b}$, $h$ and
$b$, where $b$ is a bias factor motivated by gasdynamical simulations
which suggest that the baryon fraction in clusters is slightly lower
than for the universe as a whole (\eg Eke, Navarro \& Frenk 1998;
Bialek \etal 2001). We use the results of Eke \etal (1998) from
simulations of 10 clusters of similar masses to the observed systems
to constrain $b$. Excluding the data for the most dynamically active
cluster in that study (recall that the $f_{\rm gas}$ data are drawn
from Chandra observations of dynamically relaxed systems), the
simulated clusters show consistent, relatively flat baryonic mass
fraction profiles for radii $r > 0.2 r_{\rm vir}$ (Fig~\ref{fig:eke}).
At $r=0.25 r_{\rm vir}$, a radius comparable to the measurement radius
for the Chandra observations, the simulations of Eke \etal (1998) give
$b=0.824\pm0.033$.

We note the excellent agreement between the median, scaled $f_{\rm
gas}(r)$ profile determined from the Chandra data (shown as dark
circles in Fig~\ref{fig:eke}) and the simulated profiles for the three
most relaxed clusters in the study of Eke \etal (1998; the darker
curves in Fig~\ref{fig:eke}). This agreement supports the use of the
simulations in estimating the bias factor at $r_{2500}$. Note also
that the simulations of Eke \etal (1998) indicate negligible evolution
of the bias parameter (measured within $r \sim 0.5 r_{\rm vir}$) 
over the redshift range considered here.

For our analysis of the Chandra $f_{\rm gas}$ data alone, we employ
simple Gaussian priors on $\Omega_{\rm b}h^2$ and $h$.  Two separate
sets of priors were used: `standard' priors with $\Omega_{\rm
b}h^2=0.0214\pm0.0020$ (Kirkman \etal 2003) and $h=0.72\pm0.08$
(Freedman \etal 2001), and `weak' priors in which we triple the
nominal uncertainties to give $\Omega_{\rm b}h^2=0.0214\pm0.0060$ and
$h=0.72\pm0.24$. (No priors on $\Omega_{\rm b}h^2$ and $h$ are 
assumed in the combined analysis of $f_{\rm gas}$ and CMB data; 
Section 3.3.)

We assume a Gaussian prior on $b$. The rms fractional deviation in $b$ 
from the simulations of Eke \etal (1998; $\sim 4$ per cent) was added in 
quadrature to a nominal 10 per cent systematic uncertainty associated 
with the overall normalization of the $f_{\rm gas}(z)$ curve. This allows 
for residual uncertainties associated with the simulations and/or the 
calibration of the Chandra instruments.\footnote{The agreement between 
the independent mass measurements from X-ray and gravitational lensing 
studies for several of the target clusters argues that the systematic 
uncertainties are unlikely to significantly exceed 10 per cent.} Thus, 
for the analysis of the $f_{\rm gas}$ data alone using the standard priors, 
the $\chi^2$ value for any particular model is

\begin{eqnarray}
\chi^2& =& \left( \sum_{i=1}^{26}
\frac{\left[f_{\rm gas}^{\rm SCDM}(z_{\rm i})- f_{\rm gas,\,i}
\right]^2}{\sigma_{f_{\rm gas,\,i}}^2}\nonumber \right) 
\end{eqnarray}

\begin{eqnarray}
+\left(\frac{\Omega_{\rm b}h^2-0.0214}{0.0020} \right)^2 
+\left(\frac{h-0.72} {0.08} \right)^2+\left(\frac{b-0.824} {0.089} 
\right)^2.
\end{eqnarray}
\vspace{0.1cm}

\noindent Here $f_{\rm gas,\,i}$ and $\sigma_{f_{\rm gas,\,i}}$ are
the observed values and symmetric rms errors  for the SCDM cosmology 
from Table \ref{table:fgas}. 
Note that we have not accounted for the intrinsic cluster-cluster 
scatter in $b$ explicitly. However, the simulations suggest that this 
scatter is small when compared with the statistical uncertainties in 
the $f_{\rm gas}$ measurements.

\subsection{Combination of $f_{\rm gas}$ and CMB constraints}

For the combined Chandra+CMB analysis, we importance sample 
the MCMC results from the CMB analysis, folding in the 
$f_{\rm gas}$ constraints (Allen, Schmidt \& 
Bridle 2003). Each of the MCMC samples from the CMB analysis 
provides a value for $\Omega_{\rm m}$, $\Omega_{\rm X}$, $w_{\rm X}$, 
$H_0$ and $\Omega_{\rm b}h^2$. Using these values, we fit the 
$f_{\rm gas}(z)$ data with the model described by 
Equation~\ref{eq:fgas}, including the same Gaussian prior on 
the bias factor, including the allowance for systematic 
uncertainties in the normalization of the curve. 
This provides a $\chi^2$ value for each of the 
MCMC samples. The weight of the MCMC sample is then multiplied 
by $e^{-\chi^2/2}$.

\section{Cosmological constraints}

\subsection{Results for the $\Lambda$CDM cosmology}\label{section:lcdmres}

\begin{figure}
\vspace{0.5cm}
\hbox{
\hspace{0.0cm}\psfig{figure=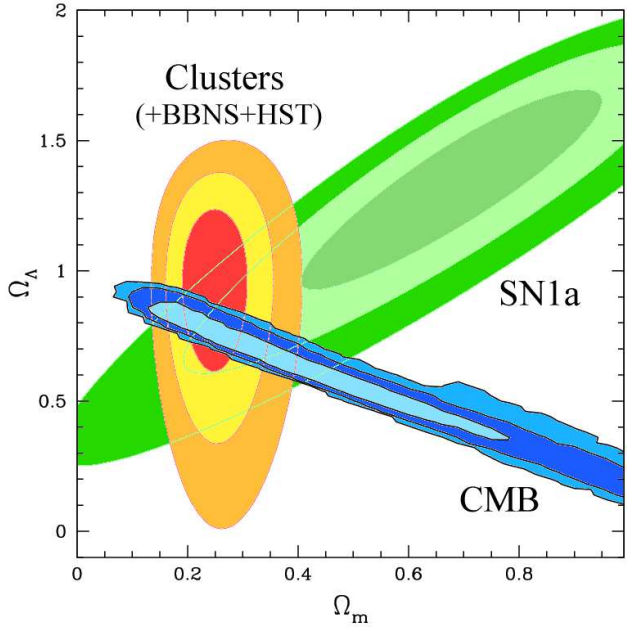,width=.49\textwidth,angle=0}
} \caption{The 68.3, 95.4 and 99.7 per cent (1, 2 and 3$\sigma$) 
confidence constraints in the $\OM,\OL$ plane obtained from the 
analysis of the cluster $f_{\rm gas}$ data using standard priors on 
$\OBH$ (Kirkman \etal 2003) and $h$ (Freedman \etal 2001). Also shown 
are the independent results obtained from CMB data using a weak uniform 
prior on $h$ ($0.3<h<1$), and Type 1a supernovae data (Tonry \etal 2003). 
A $\Lambda$CDM cosmology is assumed with the curvature, $\OK$, included 
as a free parameter in the analysis.}\label{fig:lcdm}
\end{figure}

\begin{figure}
\vspace{0.5cm}
\hbox{
\hspace{-0.5cm}\psfig{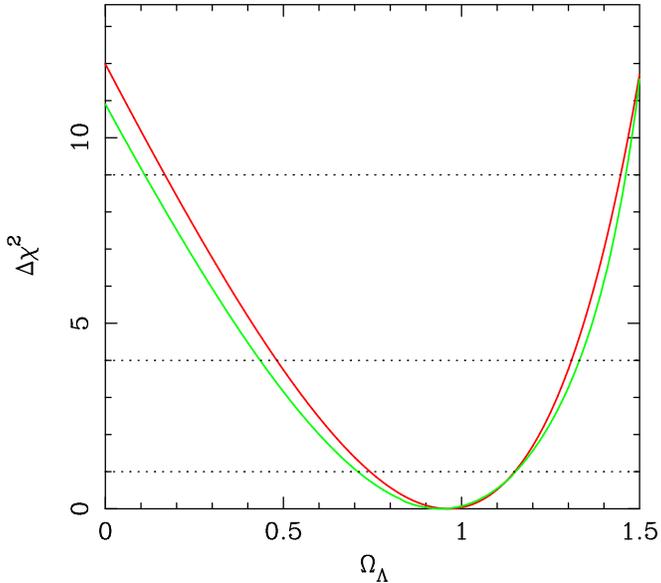}
} \caption{The marginalized constraints on $\OL$ obtained from the 
Chandra $f_{\rm gas}$ data using the standard (dark curve) and
weak (light curve) priors on $\OBH$ and $h$. The dotted lines 
mark the formal 1, 2 and 3$\sigma$ limits (see text for details). 
The curvature is a free parameter in the analysis.}\label{fig:lcdm_marg}
\end{figure}

\begin{figure}
\vspace{0.5cm}
\hbox{
\hspace{-0.5cm}\psfig{figure=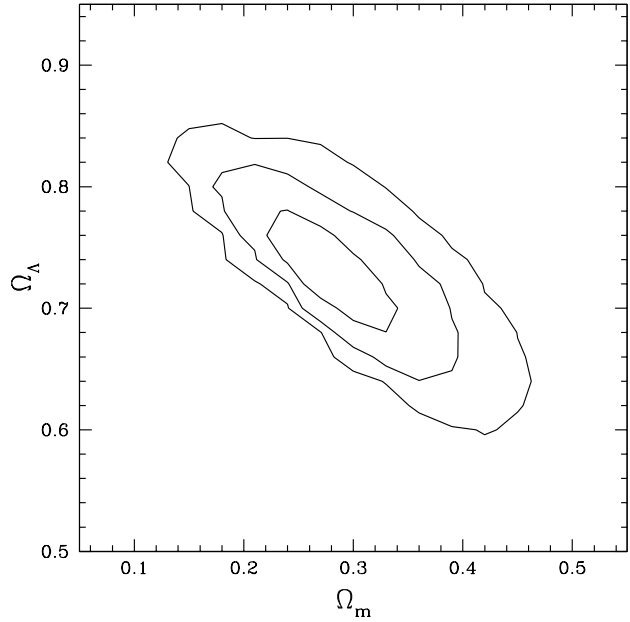,width=.49\textwidth,angle=0}
} \caption{The 68.3, 95.4 and 99.7 per cent 
confidence constraints in the $\OM,\OL$ plane obtained from the 
analysis of the combined $f_{\rm gas}$+CMB data set using the $\Lambda$CDM 
model. We find marginalized 68 per cent confidence limits of 
$\Omega_{\rm m}=0.28^{+0.05}_{-0.04}$ and
$\Omega_{\Lambda}=0.73^{+0.04}_{-0.03}$, with $\OK=-0.01\pm0.02$.}
\label{fig:fgascmb}
\end{figure}

\begin{figure}
\vspace{0.5cm}
\hbox{
\hspace{-0.5cm}\psfig{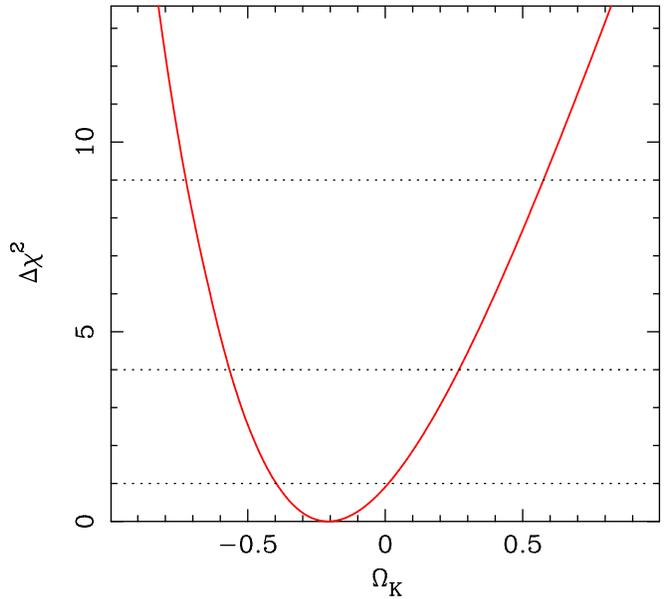}
} \caption{The marginalized constraints on $\OK$ ($=1-\OM-\OL$) obtained 
from the Chandra $f_{\rm gas}$ data using the standard priors on $\OBH$ and 
$h$. The dotted lines 
mark the formal 1, 2 and 3$\sigma$ limits (see text for details). }\label{fig:lcdm_ok}
\end{figure}

For the $\Lambda$CDM cosmology, we have examined a grid of
cosmological models covering the plane $0.0< \Omega_{\rm m} < 1.0$ and
$0.0< \Omega_{\Lambda} < 2.0$. The joint 68.3, 95.4 and 99.7 per cent
confidence contours (corresponding to $\Delta \chi^2$ values of 2.30,
6.17 and 11.8, respectively) obtained from the Chandra $f_{\rm gas}$
data, including standard priors on $\Omega_{\rm b}h^2$ (Kirkman \etal
2003) and $h$ (Freedman \etal 2001) are shown in
Fig.~\ref{fig:lcdm}. The best-fit parameters and marginalized 68 per
cent confidence limits obtained using the standard priors are
$\Omega_{\rm m} = 0.245^{+0.040}_{-0.037}$ and $\Omega_{\Lambda} =
0.96^{+0.19}_{-0.22}$, with $\chi^2=24.5$ for 24 degrees of
freedom. The $\chi^2$ value indicates that the model provides an
acceptable description of the data.

Fig.~\ref{fig:lcdm_marg} shows the marginalized constraints on
$\Omega_{\Lambda}$ obtained using both the standard and weak priors on
$\OBH$ and $h$. We see that even using the weak priors
($\OBH=0.0214\pm0.0060$, $h=0.72\pm0.24$), the $f_{\rm gas}$ data
still provide a clear detection of the effects of $\Omega_{\Lambda}$
at $> 3\sigma$ significance ($\Omega_{\Lambda} =
0.94^{+0.21}_{-0.23}$). A Monte Carlo analysis of the data indicates
that $\Omega_{\Lambda}\leq0$ is ruled out at $>99.9$ per cent
confidence.

Fig.~\ref{fig:lcdm} also shows a comparison with independent
constraints obtained from the CMB data using only a weak uniform prior
on $h$ ($0.3<h<1.0$), and from Type 1a supernovae studies (Tonry \etal
2003). The agreement between the $f_{\rm gas}$ and CMB constraints in
particular is reassuring and motivates the combined analysis of these
data sets, discussed below. 

Fig.~\ref{fig:fgascmb} shows the constraints on $\Omega_{\rm m}$ and
$\Omega_{\Lambda}$ obtained from the analysis of the combined $f_{\rm
gas}$+CMB data set. No priors, other than the constraint
$b=0.824\pm0.089$ are assumed. We see that the $f_{\rm gas}$+CMB data
set provides a remarkably tight constraint in the $\OM$,$\OL$ plane,
with best fit values and marginalized 68 per cent confidence limits of
$\Omega_{\rm m}=0.28^{+0.05}_{-0.04}$ and
$\Omega_{\Lambda}=0.73^{+0.04}_{-0.03}$. 

Fig.~\ref{fig:lcdm_ok} shows the marginalized constraints on $\OK$
($=1-\OM-\OL$) from the $f_{\rm gas}$ data using the standard priors
on $\OBH$ and $h$. The best fit result of $\OK=-0.2\pm0.2$ (68 per
cent confidence limits) is consistent with the much tighter 
constraint of $\OK=-0.01\pm0.02$ obtained from the combined
$f_{\rm gas}$+CMB data set. (The CMB data alone give 
$\OK=-0.03^{+0.04}_{-0.06}$ using only a wide uniform prior on 
the Hubble constant).

\begin{figure}
\vspace{0.5cm}
\hbox{
\hspace{-0.5cm}\psfig{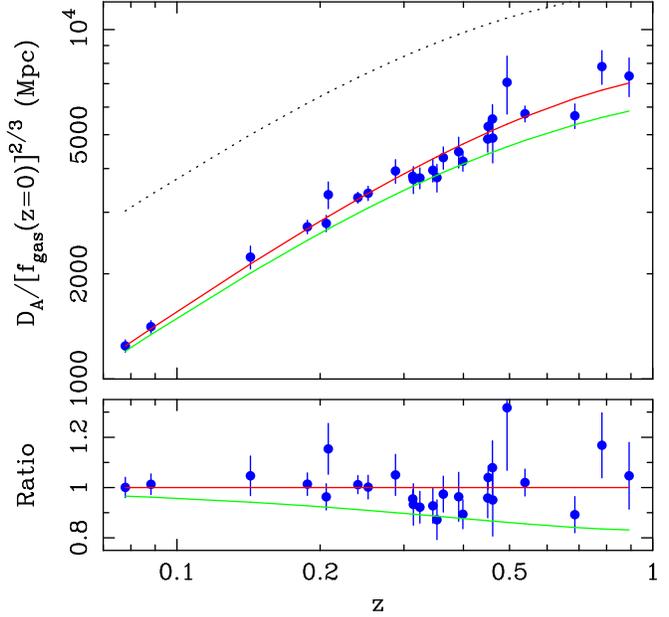}
} \caption{(Upper panel) The function $d_{\rm A}^{\rm mod}(z)/[f_{\rm
gas}^{\rm mod}(z=0)]^{2/3}$ for the best-fitting $\Lambda$CDM
cosmology (dark curve). Also shown are the curves obtained keeping all
parameters fixed at their best-fit values but setting $\OL=0$ (grey
curve), and setting $\OL=0$, $\OM=1$ (dotted curve). (Lower panel) The
ratios with respect to the best-fitting 
$\Lambda$CDM model.}\label{fig:daplot} \end{figure}

Finally, Fig.~\ref{fig:daplot} shows an alternative way to visualize the
effects of dark energy on the distances to the
clusters. The model function shown in the figure is $d_{\rm A}^{\rm
mod}(z)/f_{\rm gas}^{\rm mod}(z=0)^{2/3}$ and the data points $d_{\rm
A}^{\rm SCDM}(z)/f_{\rm gas}^{\rm obs}(z)^{2/3}$. The use of this
function removes all dependence on the reference cosmology in the
figure. The dark curve in Fig.~\ref{fig:daplot} shows the results for
the best-fitting $\Lambda$CDM cosmology. Also shown are the curves
obtained keeping the parameters fixed at their best-fit values, 
but setting $\OL=0$ (grey curve; this shows the effects of 
the dark energy component), and setting $\OL=0$, $\OM=1$ (dotted curve). 
The model including the dark energy component provides the best 
description of the data over the full redshift range. The $\OL=0$, 
$\OM=1$ model provides a poor fit, and would require $H_0\sim 20$\kmpspMpc~to 
approximately match the observed normalization (given the standard 
set of priors).

\subsection{Extended XCDM models}

\begin{figure}
\vspace{0.5cm}
\hbox{
\hspace{-0.5cm}\psfig{figure=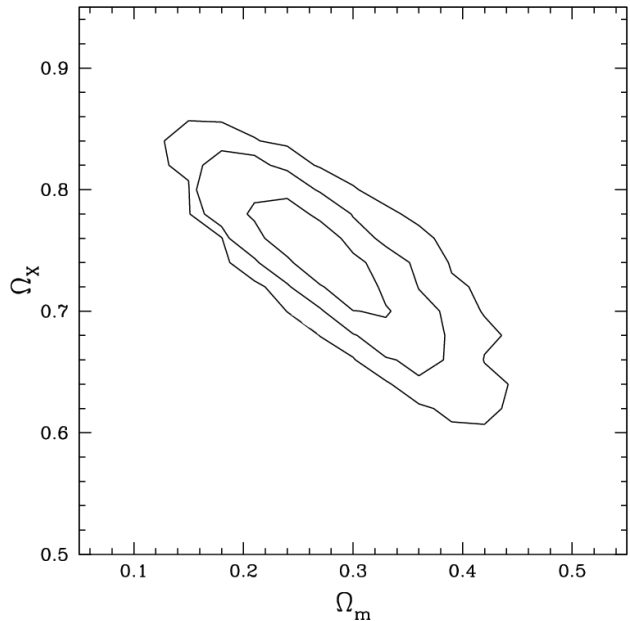,width=.49\textwidth,angle=0}
} \caption{The 68.3, 95.4 and 99.7 per cent 
confidence constraints in the $\OM,\OX$ plane obtained from the 
analysis of the combined $f_{\rm gas}$+CMB data set. We obtained
marginalized 68 per cent confidence limits of $\Omega_{\rm m}=0.26^{+0.06}_{-0.04}$ and 
$\Omega_{\rm X}=0.75\pm0.04$. The results are similar to those obtained
for the $\Lambda$CDM models in Fig.~\ref{fig:fgascmb}.
}\label{fig:xcdm}
\end{figure}

Fig.~\ref{fig:xcdm} shows the 68.3, 95.4 and 99.7 per cent 
confidence constraints in the $\OM,\OX$ plane for the extended XCDM 
models from the analysis of the combined $f_{\rm gas}$+CMB data set. 
We obtain best fitting values and marginalized 68 per cent confidence limits 
of $\OX=0.75\pm0.04$ and $\OM=0.26^{+0.06}_{-0.04}$. The constraints 
on the mean matter and dark energy densities for the 
extended XCDM models are similar to those obtained
for the $\Lambda$CDM cosmology. The curvature is measured to be 
$\OK=-0.02\pm0.02$.

\begin{figure*}
\vspace{0.5cm}
\hbox{
\hspace{0.0cm}\psfig{figure=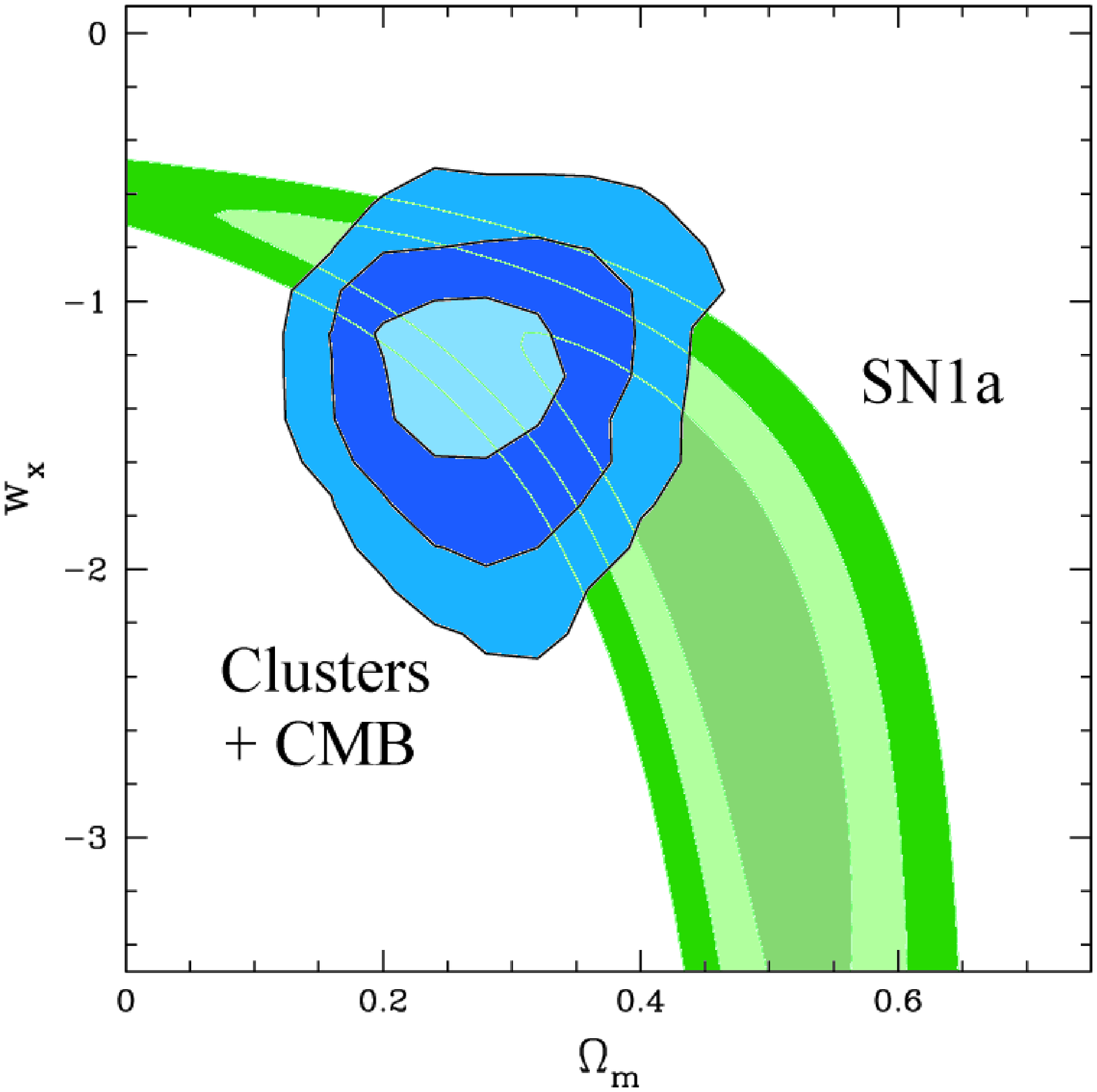,width=.49\textwidth,angle=0}
\hspace{0.5cm}\psfig{figure=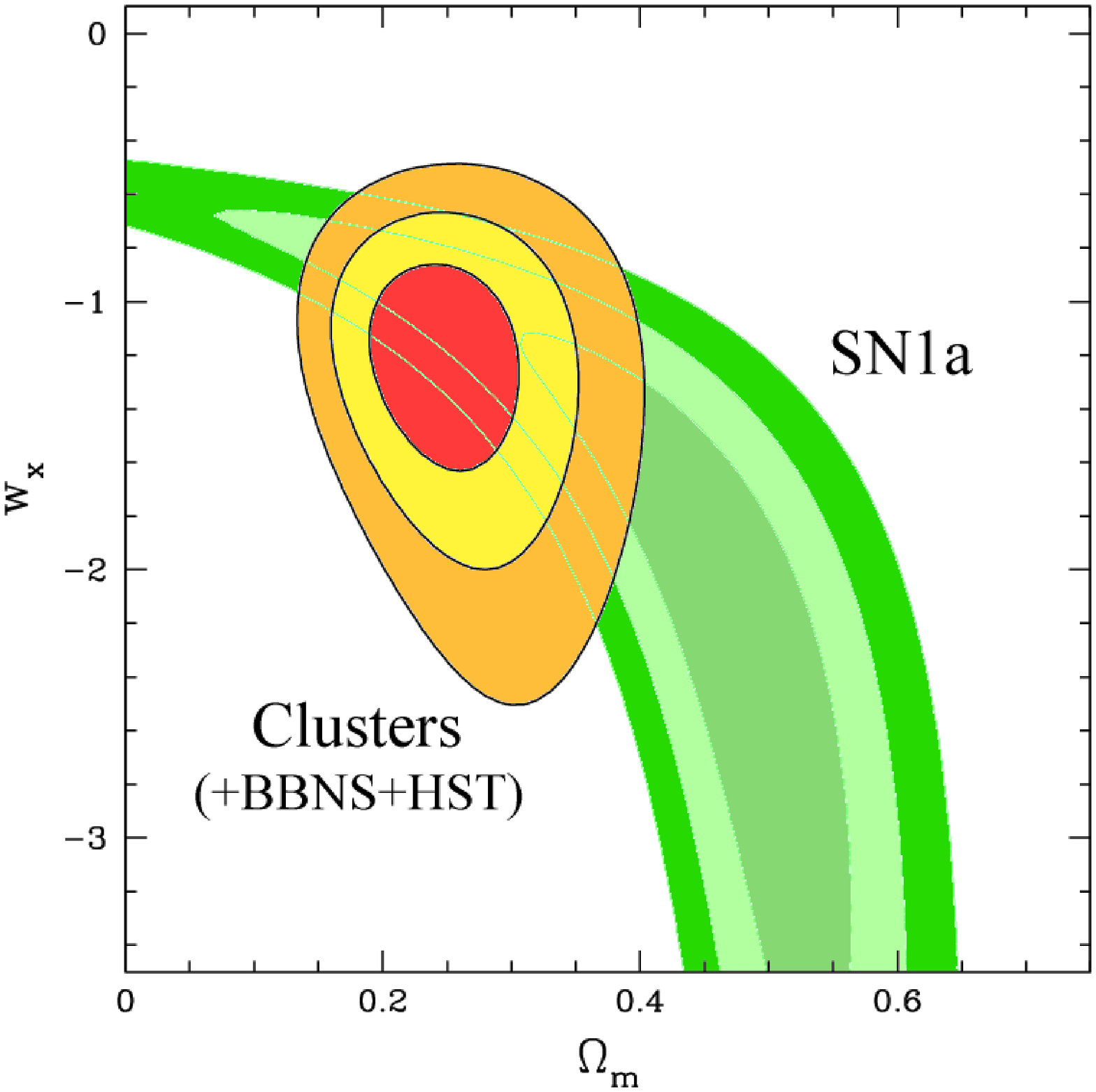,width=.49\textwidth,angle=0}
} \caption{The 68.3, 95.4 and 99.7 per cent confidence constraints in
the $\OM$,$w_{\rm x}$ plane from the analysis of (a: left panel) the
combined $f_{\rm gas}$+CMB data, with $\OK$ free.  Also shown, for
comparison purposes, are the results obtained from the analysis of
Type 1a supernovae data by Tonry \etal (2003) assuming a flat
geometry. (b: right panel) The results obtained from the Chandra
$f_{\rm gas}$ data alone, assuming standard priors on $\OBH$ and $h$
and a flat geometry, together with the supernovae constraints.
}\label{fig:omwx}
\end{figure*}

\begin{figure*}
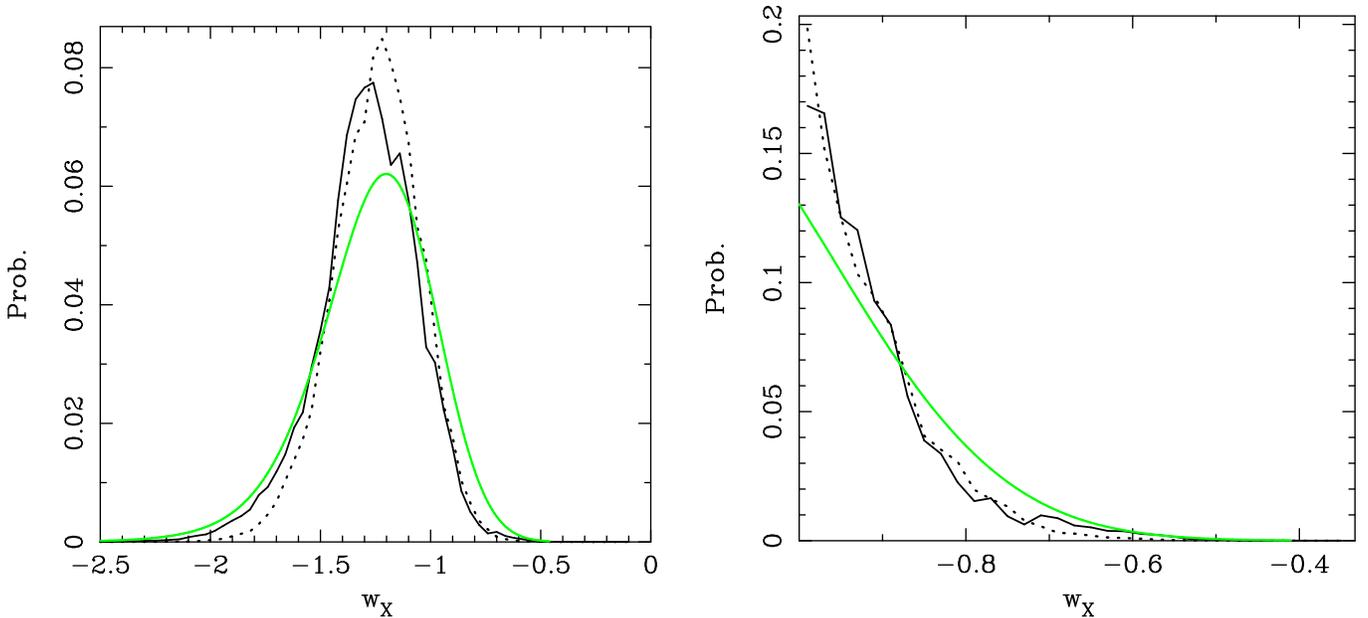

\vspace{0.5cm}
\hbox{
\hspace{0.0cm}\psfig{figure=w_marginal.ps,width=.49\textwidth,angle=270}
\hspace{0.5cm}\psfig{figure=w_marginal_restricted.ps,width=.49\textwidth,angle=270}
} \caption{(a: left panel) The marginalized constraints on the
equation of state parameter, $w_{\rm X}$, obtained from the $f_{\rm
gas}$+CMB data with $\OK$ free (dark solid
curve), $f_{\rm gas}$+CMB data assuming a flat geometry (dotted curve)
and $f_{\rm gas}$ data alone assuming a flat geometry and standard
priors on $\OBH$ and $h$ (grey curve). (b) As for (a) but imposing the
prior constraint $w_{\rm X}>-1$.}\label{fig:wx_marg}
\end{figure*}

Fig.~\ref{fig:omwx}(a) shows the constraints in the $\Omega_{\rm
m},w_{\rm X}$ plane obtained from from the same data.  Also shown, for
comparison purposes, are the results from Type 1a supernovae studies
(Tonry \etal 2003). Fig.~\ref{fig:omwx}(b) shows the results obtained
from the $f_{\rm gas}$ data alone, assuming a flat geometry and the
standard priors on $\OBH$ and $h$. The results are in excellent
agreement with those obtained from the $f_{\rm gas}$+CMB data set.
We again note the ability of the  $f_{\rm gas}$ data, used in combination
with the CMB data or standard priors, to break important degeneracies 
between parameters.

Fig.~\ref{fig:wx_marg}(a) shows the marginalized constraints on
$w_{\rm X}$ for the extended XCDM models. For the $f_{\rm gas}$+CMB
data with $\OK$ free 
we find $w_{\rm X}=-1.26^{+0.24}_{-0.24}$. Under the assumption
of a flat geometry, the same $f_{\rm gas}$+CMB data data give $w_{\rm
X}=-1.22^{+0.20}_{-0.22}$. For a flat geometry, the $f_{\rm gas}$ data
and standard priors on $\OBH$ and $h$ give $w_{\rm
X}=-1.20^{+0.24}_{-0.28}$.  Note that for a flat geometry, the
supernovae data alone give $w_{\rm X}=-2.2^{+0.8}_{-1.1}$ (68 per cent
confidence limits).

Fig.~\ref{fig:wx_marg}(b) shows the marginalized constraints on
$w_{\rm X}$ obtained from XCDM models when we apply the prior
constraint $w_{\rm X}>-1$. Under this assumption, the $f_{\rm
gas}$+CMB data with $\OK$ free give a 95.4 per cent confidence constraint of $w_{\rm
X}<-0.69$. If we assume flatness, the same data require $w_{\rm
X}<-0.75$.  For a flat geometry and standard priors on $\OBH$ and $h$,
the $f_{\rm gas}$ data alone give $w_{\rm X}<-0.69$. These constraints are
similar to those obtained by Tonry \etal (2003; $w_{\rm X}<-0.73$)
from supernovae data using a prior on $\OM$ from the 2dF Galaxy
Redshift Survey ($\OM h=0.20\pm0.03$; Percival \etal 2001) and
assuming a flat geometry. Our results on $w_{\rm X}$ are also
consistent with (and comparable to) those reported by the WMAP team
(Spergel \etal 2003).

Finally, we note that our results for the extended XCDM cosmology
imply that the mean matter and dark energy densities become equal at a
redshift $z= (\Omega_{\rm m}/\Omega_{\rm X})^{1/3w_{\rm X}}-1 =
0.30\pm0.09$, and that the Universe moves from a decelerating to an
accelerating phase at $z= [-(1+3w_{\rm X})\Omega_{\rm X}/\Omega_{\rm
m}]^{-1/3w_{\rm X}}-1=0.70\pm0.11$ (68 per cent confidence
limits).

\section{Discussion}

Our results provide the first clear confirmation of type Ia supernovae
results in terms of detecting the effects of dark energy on distance
measurements to a separate, well-defined source population. Our
results cover the redshift range where the expansion of the Universe
moves from a decelerating to an accelerating phase. The significance
of our detection of dark energy is $>3\sigma$ ($>99.9$ per cent
significance from Monte Carlo simulations) for the standard
$\Lambda$CDM model with $\OK$ free, using only weak priors on
$\Omega_{\rm b}h^2$ and $h$. This accuracy is comparable to that
obtained from current supernovae work (Tonry \etal 2003; see also
Riess \etal 2004).

It is interesting to note that our preferred value for $w_{\rm X}$ in
the extended XCDM models is slightly less than -1, which allows the
possibility that the dark energy density is increasing with time.
Such a scenario is also mildly favoured by recent Type 1a supernovae
studies (Tonry \etal 2003; Riess \etal 2004). We stress, however, that
the Chandra results remain consistent with $\Lambda$CDM ($w_{\rm
X}=-1$).

A major benefit of our technique is that the application of standard
priors on $\Omega_{\rm b}h^2$ and $H_0$, or the combination with CMB
data, also leads to tight constraints on $\Omega_{\rm m}$, thereby
allowing important degeneracies between parameters to be broken.  For
a $\Lambda$CDM cosmology ($\OK$ free), we find $\Omega_{\rm m} =
0.245^{+0.040}_{-0.037}$ using standard priors on $\Omega_{\rm b}h^2$
and $h$, or $\Omega_{\rm m} = 0.28^{+0.05}_{-0.04}$ when the $f_{\rm
gas}$ and CMB data are combined. These constraints are comparable to
those obtained from the combination of current CMB data with a variety
of other data sets and priors (\eg Spergel \etal 2003). We note that
the lower value of $\Omega_{\rm m}$ obtained in this work with respect
to ASF02 is primarily due to the inclusion of the bias factor $b$ in
the present study, together with changes in the prior on $\Omega_{\rm
b}h^2$.  The (slightly) larger error bars on $\Omega_{\rm m}$ are
due to the inclusion of the 10 per cent systematic uncertainty in the
normalization of the $f_{\rm gas}$ curve (via $b$), which is motivated by
residual uncertainties in the calibration of the Chandra detectors. It
may be that this 10 per cent allowance overestimates the systematics
errors. If it were not included, the
constraint on $\OM$ for the $\Lambda$CDM cosmology from the $f_{\rm
gas}$ data using standard priors on $\Omega_{\rm b}h^2$ and $h$ would
become $\Omega_{\rm m} = 0.246^{+0.033}_{-0.029}$.  Recall that the
constraint on $\OM$ arises primarily from the normalization of the
$f_{\rm gas}(z)$ curve and so is affected by the 10 per cent
systematic uncertainty, whereas the constraint on $\OX$ is determined
by the shape of the curve and is so largely independent of the
uncertainties in $\OBH$, $h$ and $b$.

The evidence for dark energy from the Chandra data is robust against
uncertainties in the bias factor, $b$. The results depend primarily
upon the shape of the $f_{\rm gas}(z)$ curve and so doubling the
overall uncertainty in $b$ to 20 per cent has little effect. Only
redshift evolution in $b$ can change the results on dark energy.
However, to remove the evidence for dark energy (i.e. measure $OX=0$)
we would require $b$ to decrease with increasing redshift by $>30$ per
cent over the interval $0<z<1$. This change in $b$ is much larger than
is allowed by simulations; the study of Eke \etal (1998) indicates
negligible evolution over the redshift range studied here. For
illustration purposes only, we have examined the effects of including
(substantial) evolution in $b$ such that $b(z)=(1-0.1z)b(0)$. This
leads to only a small change in the results:
$\Omega_{\Lambda}=0.72^{+0.24}_{-0.27}$ for the $\Lambda$CDM cosmology
using the Chandra data and weak priors.  (The detection of dark energy
remains significant at the $\sim 2.5$ sigma level.) Including such
evolution in $b$ also shifts the best-fit value for $w$ closer to -1 :
$w=-0.98^{+0.21}_{-0.24}$ for the same data using the XCDM model.

An important aspect of the present work is that the clusters studied
are regular, apparently dynamically relaxed systems. This results in a
significant reduction of the scatter in the $f_{\rm gas}$ measurements
with respect to studies that do not include such a selection criterion
(\eg Ettori \etal 2003). Note also that our analysis does not impose a
parametric form for the X-ray gas distribution, uses a realistic
parameterization for the total matter distribution, and makes full use
of information on the temperature profiles in the
clusters. Independent confirmation of the total masses within
$r_{2500}$ is available from weak lensing studies in a number of
cases, which lends support to the reliability of the $f_{\rm gas}$
measurements (see discussion in ASF02; a program to expand the weak
lensing measurements to the entire sample studied here is underway.)
Finally, we note that the effects of departures from spherical
symmetry on the $f_{\rm gas}$ results are expected to be small
($\approxlt$ a few per cent; Buote \& Canizares 1996, Piffaretti,
Jetzer \& Schindler 2003).

An ASCII table containing the redshift and $f_{\rm gas}(z)$ data
is available from the authors on request.

\section*{Acknowledgements}

We are grateful to V. Eke for providing detailed results from his
simulations, J. Tonry for making his supernovae data set and analysis
code available and A. Lewis for the CosmoMC code. We thank A. Edge and
E. Barrett for their ongoing efforts with the MACS, A. Vikhlinin for
advice concerning Chandra analysis, and S. Bridle and J. Weller for
discussions regarding the analysis of CMB data and comments on the
manuscript. We are grateful to C. Jones and collaborators for sharing
the latest data on RXJ1347.5-1145 and the WARPS team for allowing
early access to their Chandra data for ClJ1226.9+3332.  We thank
M. Rees for helpful comments and encouragement and R. Johnstone for
numerous discussions and help with software issues. We also thank the
anonymous referee for a helpful and rapid report. SWA and ACF thank
the Royal Society for support. HE acknowledges financial support from
grants GO1-2132X, GO2-3168X, GO2-3162X, and GO3-4164X.

\end{document}